\documentclass[pss,fleqn,embeddedheads]{w-art}

\usepackage{times}
\usepackage{w-thm}
\usepackage{amssymb}
\usepackage{cite}

\usepackage[]{graphicx}
\graphicspath{{Figures/},.}         

\begin{document}

\DOIsuffix{theDOIsuffix}
\Volume{XX}
\Issue{1}
\Copyrightissue{01}
\Month{08}
\Year{2007}
\pagespan{1}{}
\Receiveddate{\sf zzz} \Reviseddate{\sf zzz} \Accepteddate{\sf
zzz} \Dateposted{\sf zzz}

\subjclass[pacs]{07.60.Fs, 78.20.Fm, 78.20.Ci, 61.46.Fg}




\title{Optical properties of nanostructured GaSb}

\author{M. Kildemo\footnote{Corresponding
     author: {\sf Morten.Kildemo@phys.ntnu.no}}\inst{1}}
\author{I.S.\ Nerb\o\inst{1}}
\author{E. S{\o}nderg{\aa}rd\inst{2}}
\author{L. Holt\inst{1}}
\author{I. Simonsen\inst{2}}
\author{M.~Stchakovsky\inst{3}}

\address[\inst{1}]{Applied Optics Group, Department of Physics, 
            Norwegian University of Science and Technology~(NTNU),
            NO-7491 Trondheim, Norway}
\address[\inst{2}]{UMR 125 Unit\'e mixte CNRS/Saint-Gobain
        Laboratoire Surface du Verre et Interfaces,
        39 Quai Lucien Lefranc, F-93303 Aubervilliers Cedex, France}
\address[\inst{3}]{Horiba Jobin Yvon, 16-18 rue du canal, F-91165 Lonjgumeau Cedex, France}


\begin{abstract}
  Optical measurements of nanostructured GaSb prepared by sputtering
  is presented. The optical response is studied by Mueller Matrix
  Ellipsometry~(MME) in the visible range ($430$--$850 \mbox{nm}$),
  and by spectroscopic ellipsometry in the range $0.6$--$6.5
  \mbox{eV}$.  The nano-structured surfaces reported in this work,
  consist of densely packed GaSb cones approximately $50 \mbox{nm}$
  high, on bulk GaSb. The nanostructured surfaces are here shown to
  considerably modify the optical response of the surface, hence
  giving a strong sensitivity to the far field spectroscopic (Mueller
  matrix) ellipsometric measurements. The off-specular scattering and
  the depolarization is found to be low. The anisotropic response is
  particularly emphasized by studying nano-structured GaSb cones
  approximately $45$ degrees tilted with respect to the surface
  normal.  In the latter case, one observes upon rotating the sample
  around the surface normal, that the Mueller matrix elements $m_{13}$
  and $m_{14}$ oscillate as a function of the rotation angle. Finally,
  Mueller matrix techniques have been applied to the measured data, in
  order to analyze the acquired Mueller matrix in terms of physical
  realizability and noise.
\end{abstract}

\maketitle                   

\renewcommand{\leftmark}{M. Kildemo et al.: Optical properties of nanostructured GaSb}


\section{Introduction}

The field of research related to nano-structured materials, with their
wide range of applications, including photonics applications, gives
interesting problems also to the far field techniques such as {\it
  e.g.}  spectroscopic ellipsometry~(SE).  In particular, traditional
thin film properties, may be mimicked by nanostructures, and further
supply new or enhanced properties. We are here particularly focusing
on the optical response of nanostructured GaSb, as measured by MME and
generalized~SE. In this preliminary paper we primarily report the
Mueller matrix measurements and their sensitivity to
nano-structuration.

\section{Experimental}

The samples were low ion energy sputtered crystalline GaSb(001). Under
low ion energy sputtering conditions, one finds conditions where an
apparently ``smooth'' surface forms, but which in reality consists of
nano-structured cones ({\it i.e.} equivalent to a high quality
nano-structured thin film on top of the substrate). By sputtering at
normal incidence, one could obtain typically cones normal to the
sample surface (denoted ``normal cones'' in this paper), while by
sputtering at $45$ degrees incidence, with respect to the sample normal,
close to $45$ degrees tilted cones forms (denoted here the 45 degree
cones). Such issues have been studied by Atomic Force Microscopy,
Scanning Electron Microscopy (SEM), and Transmission Electron
Microscopy (TEM)~\cite{ref1,ref2,ref3}.

\begin{figure}[tbh]
   \centering
   \includegraphics*[width=0.6\linewidth,height=0.4\linewidth]{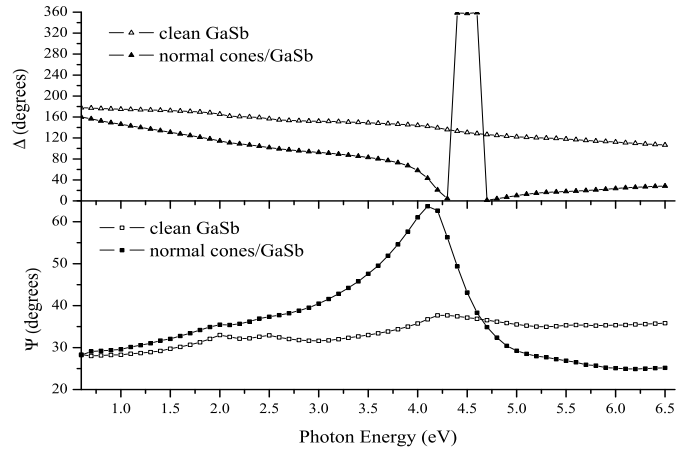}
   \label{Fig:1}
   \caption{Measured $\Psi$ and $\Delta$ from c-GaSb with
     approximately $7 \mbox{nm}$ oxide (hollow symbols), and normal
     GaSb cones on bulk GaSb (full symbols).}
\end{figure}

The optical far field measurements were performed using a commercial
Photo-Elastic-Modulator Spectroscopic Ellipsometer~(PMSE) in the range
$0.6$--$6.5 \mbox{eV}$~(UVISEL), at $55$ degrees angle of incidence.
The complete Mueller matrix was also measured using a commercial
ferroelectric liquid crystal retarder based Mueller matrix
Ellipsometer~(MM16) in the range $850$--$430 \mbox{nm}$ ($1.46$--$2.88
\mbox{eV}$), at $70^{\circ}$ angle of incidence. The sample
orientation with respect to the incoming beam was carefully recorded,
and the sample was rotated manually in steps of $45^{\circ}$, with a total
sample rotation in all cases of at least $360$ degrees.

The PMSE measurements were performed in the UVISEL setup,
polarizer-sample-PEM-analyzer, where the angle of the fast axis of the
PEM with respect to the analyzer is fixed to $45^{\circ}$.
Measurements were performed in the standard PMSE configurations
($M=0^{\circ}$, $A=45^{\circ}$), determining $I_{s}=-m_{43}=\sin
2\Psi\sin\Delta$ and $I_{c1}=m_{33}=\sin 2\Psi \cos\Delta$.  For the normal cones,
measurements were additionally performed in the configuration
($M=45^{\circ}$, $A=45^{\circ}$), determining $I_{c2}=-m_{12}=\cos 2\Psi$.
In the case of the $45$ degrees tilted cones, a complete set of $8$
measurements were performed, in order to determine the $3$ first columns
of the full Mueller matrix~\cite{ref4,ref5}.

\section{Results and discussion }

Figure~\ref{Fig:1} shows standard PMSE measurements of the
nano-structured sample for the ``normal cones''.  The ``normal cones''
were determined from high resolution-TEM to be approximately $50
\mbox{nm}$ tall. Figure~\ref{Fig:1} is of major importance, since it
demonstrate that spectroscopic ellipsometry is highly sensitive to
nanostructuration.  Furthermore, it is evident from Figure~\ref{Fig:1}
that these nanostructured surfaces strongly modify the optical
response of the system.

\begin{figure}[tbh]
   \centering
   \includegraphics*[width=0.5\linewidth,height=0.35\linewidth]{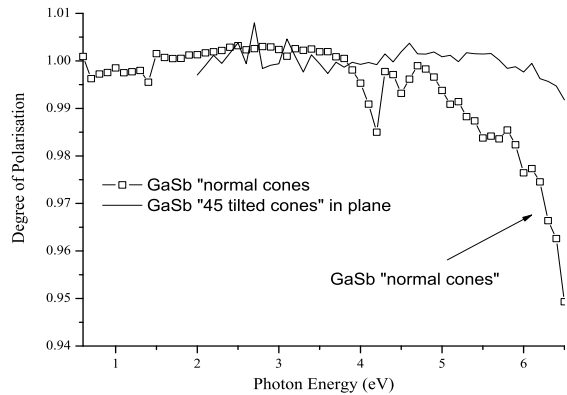}
   \label{Fig:2}
   \caption{The degree of polarization~($p$), as defined by
     equation~\protect\eqref{eq:0}, for the normal cones (hollow
     squares), $45$ degree cones (full line).}
\end{figure}

\begin{figure}[tbh]
   \centering
   \includegraphics*[width=0.95\linewidth,height=0.7\linewidth]{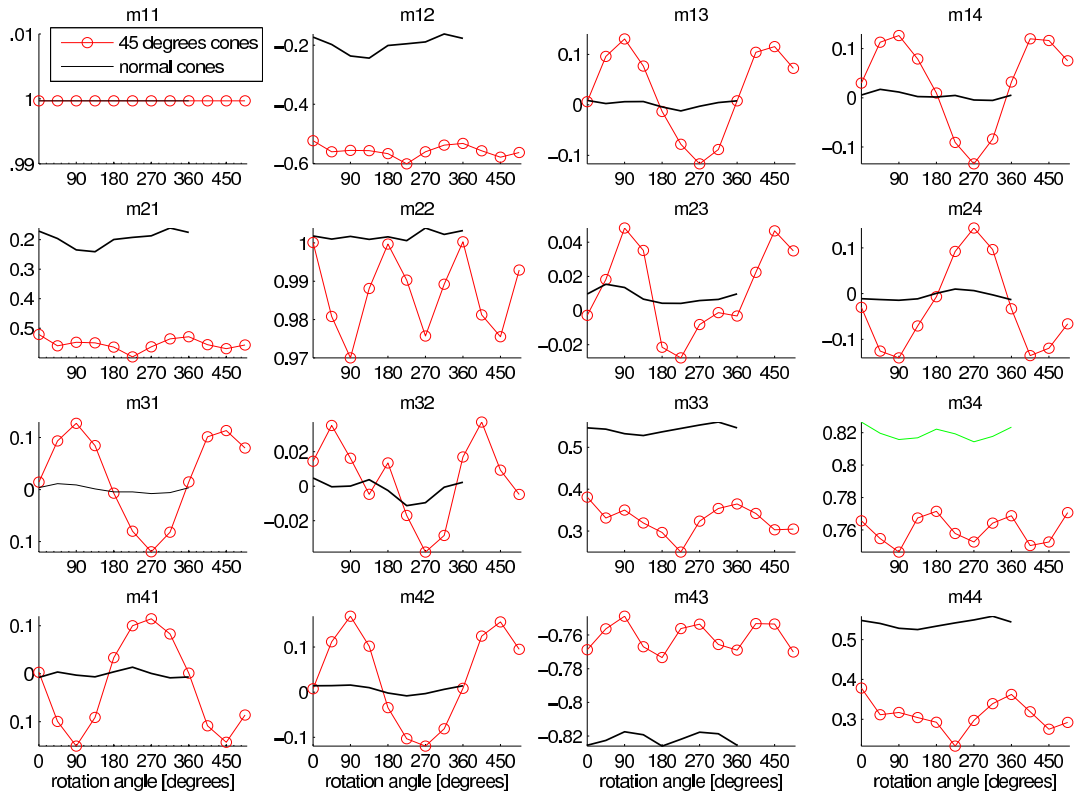}
   \label{Fig:3}
   \caption{The measured (normalized) Mueller matrix at $450
   \mbox{nm}$ ($2.755 \mbox{eV}$) as a function of sample rotation
   around the sample normal, for ``normal cones'' (full line), and 45
   degrees tilted cones (hollow circles).}
\end{figure}

The sample depolarization was found to be low, and negligible below
$3.5 \mbox{eV}$ (see Figure~\ref{Fig:2}). In fact, using He-Ne laser
light, and a goniometer, the angular scattering was measured, and it
was found to be similar to the one from c-Si. Hence, the
nano-structured surface is expected to be well modelled within an
appropriate effective medium approximation. Indeed, the low off
specular scattering, was correlated to negligible depolarisation in
the visible. The latter was confirmed by Lu-Chipman product
decomposition~\cite{ref6} of the Mueller matrix measured in the
visible. As a result of the decomposition, the total depolarisation
was determined from the diagonal elements of the depolarization
matrix. For photon energies above $4 \mbox{eV}$, an apparent trend of
depolarization was observed for the normal cones, while not observed
for the tilted cones. The ``degree of polarization'' as defined from
\begin{align}
  p &= \sqrt{ I_s^{2} + I_{c1}^{2} + I_{c2}^{2} }
  \label{eq:0}
\end{align}
is shown in Figure~\ref{Fig:2}. Equation~\eqref{eq:0} is applicable
only when the Jones matrix is diagonal, {\it i.e.} for normal cones,
or when the cones lie in the incidence plane. The degree of
polarization for the tilted cones, was thus only calculated from the
PMSE data (Mueller elements $m_{12}$, $m_{33}$ and $m_{43}$)
corresponding to the cones in the incidence plane (see also below).
Since the spacing of the cones and the typical cone sizes are much
smaller than the wavelength of light, it is speculated that the
depolarization arises rather from the statistical distribution of the
cones, and appears to follow a trend similar to Rayleigh scattering,
and such an effect is possibly more pronounced for the normal cones.
The dip in the degree of polarisation around $4 \mbox{eV}$ for the
normal cones, may possibly be a result of small ripple effects in the
``layer thickness'' (cone height) across the measurement
spot~\cite{ref4}.  This dip coincides with the photon energy where
$I_{c}=1$ and $I_{s}=0$.

\begin{figure*}[tbh]
   \centering
      \includegraphics*[width=0.98\linewidth,height=0.7\linewidth]{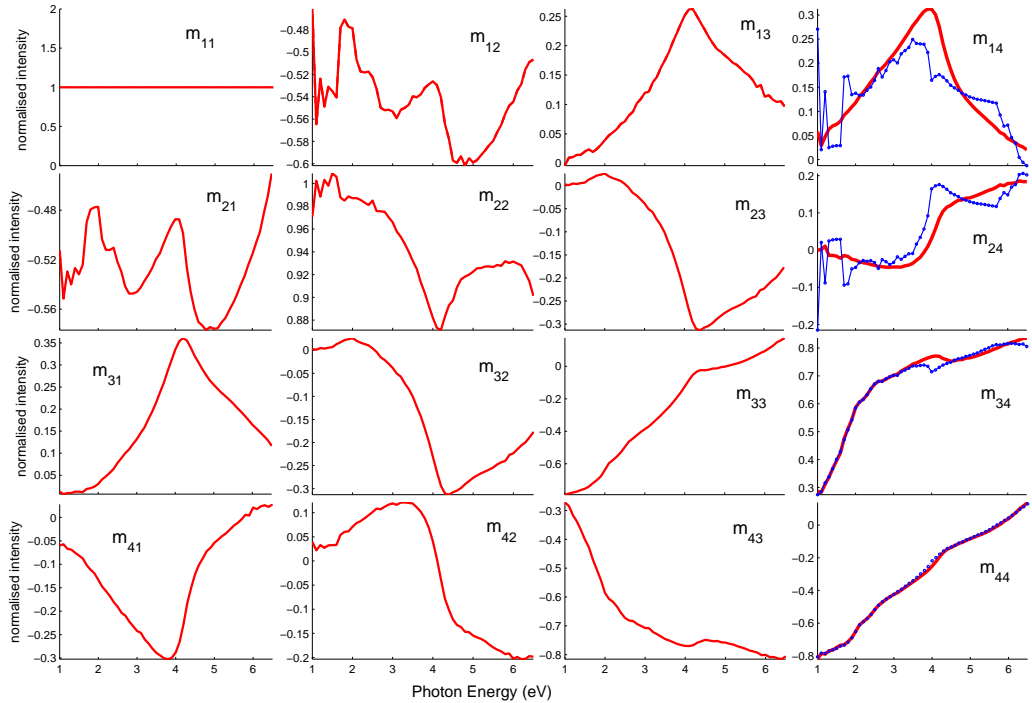}
   \label{Fig:4}
   \caption{Mueller-Jones matrix of the ``normal cones'' derived from
     the $12$ elements from generalized PMSE, at $55^\circ$ angle of
     incidence.  The last column shows the result of solving the set
     of equations as in Ref.~\citen{ref4} (hollow circles, blue),
     while the full curves~(red) are the result of the proposed method
     based on fitting the normalised elements of the Jones Matrix, and
     generating the last column of the Mueller-Jones Matrix.}
\end{figure*}

The $45$ degrees tilted cones, were both measured by complete visible
MME, and by generalized PMSE. The complete Mueller matrix at $450
\mbox{nm}$ ($2.755 \mbox{eV}$), as a function of rotation of both the
``normal cones'' and the ``tilted cones'' samples is shown in
Figure~\ref{Fig:3}. The accuracy of the manual rotation of the sample
was estimated to $\pm 5$ degrees. The movement of the ellipsometric
spot was carefully monitored to remain within a uniform sample area.
Evidently, the oval shape of the ellipsometric spot makes it
impossible to probe the exact same sample area upon rotation.The
Mueller matrix was normalized by $m_{11}$.  It is as expected,
observed from Figure~\ref{Fig:3}, that there are very little change in
the $MM$ elements upon rotation of the ``normal cones''. On the other
hand, large variations in the MM elements are observed as a function
of rotation of the sample with the ``$45$ degrees tilted cones''. In
particular, it is observed that the elements $m_{13}$ and $m_{14}$ are
quite smooth oscillating functions of the sample rotation. Both
$m_{13}$ and $m_{14}$ are zero when the cones lie in the incidence
plane, while they are maximum and anti-symmetric for the cones
pointing in a plane perpendicular to the incidence plane. Details of
the analysis and the estimated multilayer anisotropic effective medium
models and fits, in direct comparison to SEM data, will be reported
elsewhere~\cite{ref3}.

Generalized PMSE was also used to characterize the ``$45$ degrees
tilted cones'', in the range $0.6$--$6.5 \mbox{eV}$. A set of
sufficient configurations were recorded, as a function of the sample
being rotated around the sample normal, thus allowing the three first
columns of the Mueller matrix to be determined for each angle of
rotation. The normalised Mueller matrix for a general non-depolarising
anisotropic sample may be written as:
\begin{align}
  \label{eq:1}
  m_{ij}^J &= \frac{1}{2} \, \mbox{Tr} \left(J^\dagger \sigma_i J \sigma_j\right),
\end{align}
where $\sigma_{i}$ are the Pauli matrices here defined as in the work
by Cloude~\cite{ref7}, and $J$ is the normalised Jones
matrix~\cite{ref8}. This matrix is described by the complex
reflections coefficients:
\begin{align}
  \label{eq:2}
  J &= \frac{1}{r_{ss}} 
         \left( 
           \begin{array}{cc} 
               r_{pp} & r_{ps} \\ r_{sp} & r_{ss}
           \end{array} 
        \right)
    = 
     \left( 
         \begin{array}{cc} 
           \gamma_{pp}e^{i\delta_{pp}} & \gamma_{ps}e^{i\delta_{ps}}
           \\ 
           \gamma_{sp}e^{i\delta_{sp}} & 1
         \end{array} 
     \right),
\end{align}
where $\gamma_{pp}=\left|r_{pp}\right|/\left|r_{ss}\right|$, and
$\delta_{pp}$ is the complex argument of $r_{pp}/r_{ss}$. The other
quantities are defined in an equal manner.  The complete normalised
Mueller-Jones matrix only depends on these six parameters, whereas the
missing elements of the measured Mueller matrix can then be found by
solving a set of equations, as described by Jellison and
Modine~\cite{ref4}.  The last column of the Mueller matrix in
Figure~\ref{Fig:4}, shows the result of the calculated Mueller-Jones
matrix, using the latter approach (hollow circles). It is particularly
observed that the elements $m_{14}$ and $m_{24}$ are noisy and not
well determined. The method of solving the set of equations appears
thus in some cases to give unstable solutions (probably due to poor
conditioning). An alternative approach that may better handle noise
and small amounts of depolarisation has been developed. It is here
proposed to fit the normalized complex Jones elements as a function of
wavelength, to the measured $3$ first columns of the Mueller matrix, by
minimization of:
\begin{align}
  \label{eq:3}
  \alpha^2  &=
     \sum_{i=1}^4 \sum_{j=1}^3 \left( m_{ij}^{\mbox{exp}} - m_{ij}^J \right)^2,
\end{align}
where the Mueller-Jones elements $m^{J}_{ij}$ are calculated from the
fitted elements of $J$, according to eqn.~\eqref{eq:1}.
Figure~\ref{Fig:4} also shows the latter estimation of the
Mueller-Jones matrix.  It is observed from the last column in
Figure~\ref{Fig:4}, that this procedure appears more numerically
stable than the direct solution from the set of equations. It is
further observed from Figure~\ref{Fig:4}, that a particular enhanced
sensitivity to the nano-structuring of GaSb is found around $4
\mbox{eV}$. The Mueller matrices measured from both the visible-MME
and estimated from generalized SE, were tested for their physical
realizability, using the method described by Cloude and
Pottier~\cite{ref9}.  The entropy and physical realisability of the
Mueller matrices, as defined by Cloude {\it et al.}~\cite{ref9}, was
used to determine any possible depolarization effects.  For the
visible MME measurements in the limited range $1.46$--$2.88
\mbox{eV}$, the entropy was found to be particularly low, and the
physical realisability good. Low entropy indicates no depolarization,
while the physical realisability indicates little measurement noise.
This was also the case for the PMSE Mueller-Jones matrices in the
energy range from $1$ to $2.5 \mbox{eV}$. In the range $2.5 \mbox{eV}$
to $6.5 \mbox{eV}$, the physical realisability was within reasonable
limits, {\it i.e.}  accepting the matrices as proper Mueller matrices
that allow detailed analysis, but a higher noise level might hide
minor depolarizing effects.

\section{Conclusion} 

Spectroscopic ellipsometry and generally Mueller matrix ellipsometry
have been shown to be a useful technique for the characterization of
nanostructured surfaces such as nanocones of GaSb on GaSb.


\end{document}